\pdfoutput=1
\documentclass[runningheads]{llncs}
\usepackage{graphicx}
\usepackage{algorithm}
\usepackage{algpseudocode}
\usepackage{amsmath}
\usepackage{amssymb}
\usepackage{hyperref}
\usepackage{xcolor}

\algrenewcommand{\algorithmiccomment}[1]{\hfill$\triangleright$ #1}

\begin{document}
\title{Verifiable Computation with Trusted Execution Environments and On-Chain Digital Rights Tokens}
\titlerunning{Verifiable computation with TEEs and On-Chain DRTs }
%
\author{Bingle Stegmann Kruger\inst{1,2} \and
Co-Pierre Georg\inst{1}}
\authorrunning{B.S. Kruger and C.P. Georg}
%
\institute{Frankfurt School of Finance \& Management, 60322 Frankfurt am Main, Germany \and
University of Cape Town, Cape Town, South Africa
\email{krgbin001@myuct.ac.za}
}
\maketitle              
\begin{abstract}
We present an architecture that enables data owners to combine private data into data pools using \emph{Trusted Execution Environments} (TEEs) and manage these pools by issuing narrowly scoped computational rights, encoded as \emph{Digital Rights Tokens} (DRTs), to third-party data analysts. Each DRT binds specific open-source code to a pool and is issued and redeemed on a distributed ledger. Data analysts can obtain the right to execute open-source code on the combined sealed data inside a TEE and receive the result from this code execution, but not the underlying data. We argue for a \emph{control-centric} view of privacy in which creators retain ex ante control over how their data is processed. A reference implementation runs \emph{WebAssembly} (WASM)/Python jobs over sealed datasets and records redemptions on Solana, illustrating the feasibility and limitations of the platform.


\end{abstract}

\section{Introduction}

Data is a strategic asset with significant economic value~\cite{McKinsey2020OpenData}. Yet, contemporary infrastructures reward the entities that \emph{hold} data rather than the individuals and organisations that \emph{create} it~\cite{sadowski2019data}. Access-based sharing typically requires providers to disclose raw datasets, after which recipients can copy, combine, or repurpose the data beyond the intent of the original agreement. This mismatch between economic value and technical control underlies recurring privacy incidents~\cite{cheng2017enterprise,Hinds2020PrivacyConcerns,wheatley2016extreme} and constrains research and innovation that depend on sensitive datasets~\cite{Goldfarb2012}. 
One approach to this governance challenge is information-centric networking~\cite{Ahlgren2012}. Building on this line of research, Laoutaris and Iordanou~\cite{LaoutarisIordanou2021} propose combining information-centric networking with \emph{Trusted Execution Environments} (TEEs) and digital watermarking to let users regulate access to their data, receive compensation for authorised use, and foster ownership over digital information. They identify several challenges--two of which are particularly central--that any system aspiring to restore user control must overcome. First, such a system must enable the controlled sharing of data with third parties. Second, it must ensure that users are compensated for the use of their data.

In this paper, we present an information infrastructure that not only solves both challenges but also satisfies a stricter control condition. An ideal information infrastructure would enable users not only to make their data {\em available} to third parties---who could then copy the data and use it in unauthorised ways---but also to control how third parties process their data. Digital watermarking can help to trace how third parties have used data and when it has been copied, but it does not prevent the unauthorised use of the data, and in many instances, one-time unauthorised use suffices to cause harm to the users who created the data. We propose a system that allows users to create personal data vaults, combine their data with the data of other users in joint data pools, and grant third parties the right to execute code over the joint data pools without losing control over their data in the sense of Definition \ref{Def::InformationCoOwnership}. 

\subsection{Digital Privacy and Digital Rights}

For legal scholars, privacy is a contested concept with different, overlapping, competing schools of thought~\cite{Solove2008}. These include: (i) the right to be let alone~\cite{WarrenBrandeis1890}, (ii) limited access to the self~\cite{Gavison1980}, (iii) secrecy~\cite{Posner1983}, and (iv) control over personal information~\cite{Westin1967}. For this paper, we focus on \emph{digital privacy},\footnote{Various cryptographic approaches exist to provide users with a basic notion of digital privacy, the simplest of which is for the user to encrypt their data before storing it.} i.e., privacy in the context of {\em ``social and economic activity conducted online''} as defined by Fainmesser, Galeotti, and Momot~\cite{FainmesserGaleottiMomot2022}. 

We seek a suitable definition of digital privacy, without binding ourselves to the constraints of existing information systems. Towards this end, first note that all four notions of privacy contain an element of privacy-as-control over information, which motivates a control-based understanding. Westin~\cite{Westin1967} defines privacy as {\em ``the claim of individuals, groups, or institutions to determine for themselves when, how, and to what extent information about them is communicated to others.''} In Westin's view, control over information is an expression of ownership: {\em ``personal information, thought of as the right of decision over one's private personality, should be defined as a property right''}~\cite{Westin1967}. 
We adopt a notion of ownership as {\em control over a bundle of rights} in the sense of Demsetz \cite{Demsetz1967}. Taken together, this leads to the following definition:

\begin{definition}\label{Def::InformationOwnership}
Information ownership is the control over a bundle of rights attached to an information.
\end{definition}

Since data is the physical embodiment of information, this definition also naturally translates to data ownership. This formulation also addresses Westin's~\cite{Westin1967} and Solove's~\cite{Solove2008} critique that creators cannot control downstream use by allowing each right in the bundle to be specified explicitly (e.g., by binding permitted code), yet Definition~\ref{Def::InformationOwnership} still overlooks the relational nature of data~\cite{Acemoglu2022}, motivating an alternative definition in which users jointly control the bundle of rights:

\begin{definition}\label{Def::InformationCoOwnership}
Information co-ownership is the joint control over a bundle of rights attached to an information.
\end{definition}

The definition says nothing about how joint control is organised. Consequently, any system that allows users to own their data must provide a governance system within which joint control can be organised. With all of the above, we are now able to define digital privacy for this paper:

\begin{definition}
Digital privacy is a user's control over information they created.
\end{definition}


Before outlining an information infrastructure that respects users' digital privacy, it is necessary to specify how {\em ``rights attached to an information''} can be encoded. To this end, consider the rights granted to the creator of an information in a specific setting, namely when storing information in a file on a computer with a Unix-based operating system~\cite{RitchieThompson1974}. Initially, Unix provided three file rights, read ($r$), write ($w$), and execute ($x$), to three categories of users: the file's owner, members of the owner's group, and all other users. Assuming that $user_1$ created a file, the implied rights can be expressed as:
\begin{eqnarray*}
    user_1  & ::  & r_1  w_1  x_1  \\
    user_2  & ::  & r_2  w_2  x_2  \\
    \vdots & \vdots & \vdots \\
    user_N  & ::  & r_N  w_N  x_N ,
\end{eqnarray*} 
where $r_i, w_i, x_i \in \{0,1\}$ indicate the presence (1) or absence (0) of the right.
Each $user_i$, therefore, either has or does not have the right to read, write or execute a file, i.e., only up to three rights. Consequently, each user only has a minimal bundle of rights and only limited options to control these rights (e.g., transfer them to another user). On Unix, the operating system provides some protection for users' rights, with the exception of root, who can always override the file permissions. While a discussion about the legal status and nature of digital rights is beyond the scope of this paper, our goal is to create an information infrastructure that recognises digital rights as inalienable rights. 

The Universal Declaration of Human Rights~\cite{UNGA1948} recognises privacy as a human right in Article 12 and the right to own property in Article 17. If taken seriously, this implies two criteria that an information infrastructure should satisfy: First, the owner or group of owners of an information must be able to define which digital rights they hold. Consequently, they must also be able to decide which rights to grant to others. And second, the owner or group of owners of an information must not be deprived of their rights. To satisfy the first criterion, consider an information infrastructure that allows rights ($\mathtt{r}$) of the type:
\begin{eqnarray*} 
    user_1  & ::  & \mathtt{r}^1_1  \ldots \mathtt{r}^K_1  \\
    user_2  & ::  & \mathtt{r}^1_2  \ldots \mathtt{r}^K_2  \\
    \vdots & \vdots & \vdots \\
    user_N  & ::  & \mathtt{r}^1_N  \ldots \mathtt{r}^K_N ,
\end{eqnarray*} 
where $K$ is the total number of rights implemented by the infrastructure and $\mathtt{r}^k_i \in \{0,1\}$ indicates whether $user_i$ holds right~$k$ (1) or not (0).

As the number of possible rights is very large, we have decided to represent a right as a \emph{Digital Rights Token} (DRT). A DRT abstracts an individual right into a tokenised form that can be issued, transferred, and redeemed within the infrastructure. A DRT is defined by a unique name, a short description, a unique data pool address to which the DRT refers, and a unique reference to a piece of code which implements the right. DRTs also include the token supply and the unique address of a smart contract that manages the tokens. Specifically, the smart contract manages the issuance and ownership of tokens and their distribution in the primary market, as well as their redemption. Definition~\ref{Def::DRT} summarises the structure of a DRT.

\begin{definition}\label{Def::DRT}
A Digital Rights Token (DRT) is an on-chain digital asset representing specific, verifiable rights concerning a data pool. Each DRT corresponds to one unit of a specific right. A DRT has the following properties:
\begin{itemize}
  \item \texttt{drtType} : $\texttt{"append"} \text{ (right), or } \texttt{"wasm"} \mid \texttt{"python"}$ (execution runtime)
  \item \texttt{mintAddress} : $\mathrm{Address}$ (a Base58-encoded address)
  \item \texttt{managingPoolAddress} : $\mathrm{Address}$ (pool account address)
  \item \texttt{poolName} : $\mathrm{String}$ (human-readable dataset/pool name)
  \item \texttt{supply}, \texttt{cost} : $\mathrm{N}$ (nonnegative integers for token supply and cost)
  \item \texttt{githubUrl} : $\mathrm{URL}$
  \item \texttt{codeHash} : $\mathrm{Hash}$ (cryptographic hash of approved code)
\end{itemize}
\end{definition}

\paragraph{Privacy as control over use.}
Our design targets \emph{control-centric privacy} where creators decide \emph{how} their data may be processed (the allowed computations, encoded in the DRTs), \emph{by whom} (DRT holder), and \emph{under which terms} (price, quotas, and output constraints). Accordingly, data creators retain ex ante control over how their data is processed rather than who may copy it. This notion recognises that outputs may leak bounded information--by design--and makes that trade-off explicit and enforceable through mechanisms and implementation rather than policy statements alone. In contrast to watermarking or contractual restrictions, the permitted computation and its runtime environment are bound to on-chain state and an attested TEE.

\paragraph{Contributions.} In this paper, we make the following contributions:

\begin{itemize}
  \item We formalise digital rights as a $K$-dimensional rights vector and instantiate each right as a \emph{Digital Rights Token} (DRT) that encodes \emph{per-computation} licences over a specific sealed data pool, including supply, price, code reference, and a code hash. This enables the creation of secondary markets while ensuring that no data is exposed beyond what is explicitly permitted by the authorised computations.
  \item  We present an information infrastructure with a redeem-to-execute path that binds a redeemed DRT to a specific, auditable program (open-source code with an expected hash value) and an attested TEE. The DRT's on-chain immutability and an oracle trigger ensure that only authorised code executes, while RA-TLS channels and per-pool sealing keys ensure raw data never leaves the enclave and only permitted results are returned.
  \item We provide a proof-of-concept open-source implementation demonstrating end-to-end feasibility: DRT issuance and redemption on Solana and enclave execution of WASM/Python jobs over sealed pools. While not yet complete, the implementation highlights both the practicality of our approach and current limitations (e.g., oracle trust, side-channel risks, and output leakage).
\end{itemize}

\subsection{Scope and limitations}

We prioritise \emph{control} over secrecy as outputs may reveal bounded information determined by the permitted computation, and we make this trade-off explicit. Our implementation is minimal and should be read as a proof of concept rather than a complete system. The implementation currently supports CPU-based TEEs and enclaved execution of basic WASM/Python programs, but does not include multi-oracle consensus, GPU TEEs, or advanced output-privacy mechanisms beyond limiting the number of DRTs (e.g., rate limiting, aggregation-only schemas, or differential-privacy wrappers). Runtime failures of redeemed DRTs are not yet handled gracefully. In addition, known limitations of Intel SGX, such as side-channel vulnerabilities, apply to our prototype. Our contribution demonstrates how on-chain digital rights, enclave attestation, and tokenised incentives can be combined into a cohesive mechanism for control-centric data sharing.

\subsubsection{Future work.}

In this paper, we focus on presenting an information infrastructure using DRTs for verifiable computation. We leave a systematic performance evaluation of our implementation to future work, including execution overhead and the scalability of DRT issuance on-chain. Other avenues for exploration include extending support to GPU-enabled TEEs, integrating stronger output-privacy mechanisms (e.g., differential privacy or aggregation-only schemas), incorporating multiple oracle nodes, and analysing incentive and governance models for data contributors and data analysts.

\section{System Overview}

Designing an alternative information infrastructure for information co-ownership must address established dimensions of information infrastructures. Hanseth and Monteiro~\cite{HansethMonteiro2000} identify six aspects of information infrastructures. Information infrastructures (i) have an enabling function, (ii) are shared by a collection of users and user groups, (iii) are open, (iv) are socio-technical systems, (v) are connected and interrelated ecologies of networks, and (vi) develop through extending the installed base. 
We propose an alternative information infrastructure that satisfies these six aspects by enabling users to own information as per Definition~\ref{Def::InformationCoOwnership}, where co-ownership entails joint control over the rights attached to an information. 
Consequently, our system inherently aligns with the first aspect of information infrastructures.

\begin{figure}[t]
    \centering
	\includegraphics[width=0.95\textwidth,alt={Architecture diagram of the proposed information infrastructure, comprising five components. Firstly, a frontend application with a Blockchain SDK to manage pools, and buy and redeem DRTs. The second component is the enclave deployment service, which deploys new enclaves. Each of these are called execution enclaves, one per pool (Enclave 1 to Enclave N). The fourth component is the smart contract and blockchain network. The last component is, Oracle nodes that relay blockchain events such as DRT redemptions to the execution enclaves.}]{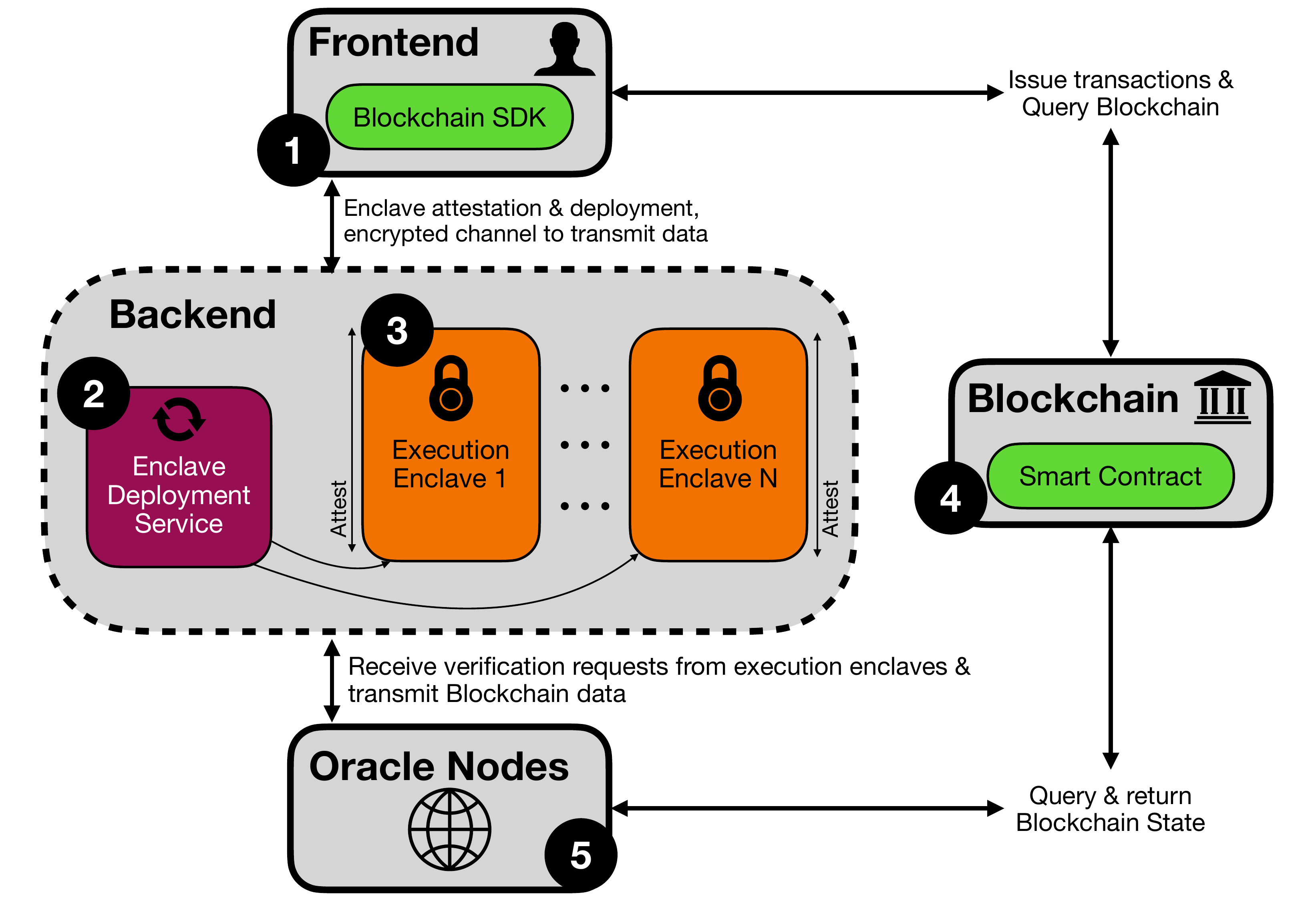}
	\caption{Architecture of the proposed information infrastructure that facilitates information co-ownership.}
    \label{Fig::SystemsDiagram}
\end{figure}

Our system\footnote{GNU AGPL v3; Zenodo: \url{https://doi.org/10.5281/zenodo.21874955} (core library), \url{https://doi.org/10.5281/zenodo.21874949} (enclave deployment service).} consists of five key components, shown in Figure~\ref{Fig::SystemsDiagram}. First, the frontend application provides users with an interface to create pools, issue and redeem DRTs, and manage contributions. Second, the enclave deployment service provisions and manages new enclaves for each data pool, maintaining isolation across pools. Third, execution enclaves, implemented using TEEs based on Intel SGX~\cite{Anati2013,Costan2016IntelSGX}, manage each data pool by sealing and unsealing\footnote{Encryption and decryption using an enclave-specific key.} data  (data always remains in the enclave), verifying DRT redemptions via an oracle, and executing only the computations authorised by the DRT. These (attestable) enclaves essentially function as an API facilitating secure computation and controlled data exchange between components. Fourth, the rights attached to each dataset (i.e., data pool) are represented as Solana SPL tokens and recorded on the Solana testnet. Solana was chosen due to its low fees, smart-contract flexibility, and configurable crypto-assets, although other permissionless blockchains with smart contract capabilities could be used with minor implications for our system. Fifth, oracle nodes bridge the blockchain and the execution enclaves by monitoring token redemptions and conveying blockchain state updates, ensuring that only authorised computations are executed inside the TEE. Together, these components enforce co-ownership by coupling tokenised rights (DRTs) with verifiable computation inside attested enclaves.

\subsection{User Roles and Workflows}

The system defines four primary roles, which are not mutually exclusive: data creators, code creators, data contributors, and data analysts. A fifth role, traders, may emerge where DRTs are exchanged for profit. Data creators define a schema (e.g., JSON) and instantiate a pool by specifying permissible DRTs and their configurations (e.g., supply and cost). A data pool (Section~\ref{sec:pools}) consists of a data package and an associated smart contract that manages the issuance, distribution, ownership and redemption of DRTs. The enclave deployment service deploys a new enclave with our library, while the associated smart contract mints the DRTs that govern the pool. Code creators provide the open-source code referenced by these DRTs. Data contributors join existing pools by purchasing and redeeming an $\mathsf{AppendDRT}$, uploading schema-conformant data to the enclave over an encrypted channel, and receiving contributor tokens that track pool ownership and revenue shares. Data analysts obtain computational access by purchasing and redeeming an $\mathsf{ExecuteDRT}$, each binding a single open-source program to a pool. The redemption of DRTs is recorded on-chain and relayed by oracles to the enclave, which unseals the data, verifies the code hash, and executes only the permitted computation. Results are returned over an encrypted channel to the analyst, while raw data remains sealed. Contributor royalties are distributed via smart contracts in proportion to ownership tokens, enabling revenue sharing without disclosure of underlying data.

\subsubsection{Data Lifecycle.}\label{sec:lifecycle}

To make the role of each component concrete, we trace data through the system from creation to result delivery. The lifecycle has four phases: \emph{(P1)} Pool creation, \emph{(P2)} Contribution, \emph{(P3)} Redemption and execution, and \emph{(P4)} Settlement.
During \emph{P1} (Algorithm~\ref{alg:pool-creation}), the data creator defines a schema and the permitted computations for the pool (i.e., the DRTs). Each permitted computation is admitted to an allow list as a \textit{(code reference, code hash)} pair and is associated with an $\mathsf{ExecuteDRT}$.\footnote{$\mathsf{AppendDRT}$ is native to the SDK and therefore has no code reference or hash.}
The smart contract registers the pool and its DRT mints. The deployment service launches a new enclave, and the creator optionally verifies the enclave through remote attestation before provisioning encrypted seed data.

\begin{algorithm}[h]
\caption{P1: Pool creation}
\label{alg:pool-creation}
\begin{algorithmic}[1]
\Require Data creator $C$, Solana program $S$, Enclave $E$,
Deployment service $D$, Pool schema $\sigma$, Code references $r_j$, Code hashes $h_j$

\State $C$ defines schema $\sigma$, permitted computations (DRTs), and
       pricing
\State $C$ publishes permitted computations as allow-listed pairs
       $(r_j,h_j)$
\State $S$ registers the pool and creates the corresponding DRT mints
\State $D$ provisions a new enclave $E$ for the pool
\State $E$ publishes an attestation quote binding its code and
       configuration
\State $C$ optionally verifies the attestation and provisions encrypted seed data to $E$
\end{algorithmic}
\end{algorithm}

For \emph{P2} (Algorithm~\ref{alg:contribution}), a contributor buys an $\mathsf{AppendDRT}$ and redeems it for the target pool. The smart contract burns the token and emits a redemption event, which the oracle relays to the enclave. The contributor uploads a record over an encrypted channel. The enclave validates the record against the pool schema and seals it. This binds the record to the attested state, so other enclave images, including downgraded ones, cannot unseal it. The contract records the contributor's pool share for later royalty distribution.
 
\begin{algorithm}[h]
\caption{P2: Contribution}
\label{alg:contribution}
\begin{algorithmic}[1]
\Require Contributor $U_i$, Solana program $S$, Oracle $O$, Enclave $E$,
Pool schema $\sigma$, DRT type $\mathsf{AppendDRT}$, Encrypted record $x_i$

\Procedure{Contribute}{$U_i,x_i$}
  \State $U_i$ redeems an $\mathsf{AppendDRT}$ for the target pool
  \State $S$ burns the token and emits a redemption event
  \State $O$ relays the redemption event to $E$
  \State $U_i$ sends encrypted record $x_i$ to $E$ over the attested channel
  \State $E$ validates $x_i$ against $\sigma$ and seals $x_i$ into the pool state
  \State $S$ records $U_i$'s contribution share
\EndProcedure
\end{algorithmic}
\end{algorithm}

Under \emph{P3} (Algorithm~\ref{alg:execution}), an analyst buys an $\mathsf{ExecuteDRT}$ bound to a specific \textit{(code reference, code hash)} pair from the pool's allow list and redeems it against the smart contract. The oracle resolves the code reference and forwards the candidate computation to the enclave. The enclave recomputes the code hash, checks it against the on-chain record, and verifies that the computation is approved for the pool. After verification, the enclave unseals the pool, executes the computation, and enforces the output policy\footnote{The declared output shape or disclosure constraints are described in Section~\ref{sec:governance}.} associated with the DRT. Raw records do not leave the protected boundary at any step and only the policy-approved (DRT) result is released. 

\begin{algorithm}[h]
\caption{P3: Redemption and execution}
\label{alg:execution}
\begin{algorithmic}[1]
\Require Analyst $A$, Solana program $S$, Oracle $O$, Enclave $E$, DRT type $\mathsf{ExecuteDRT}$, Code references $r_j$, Code hashes $h_j$,
Computation $j$

\Procedure{ExecuteJob}{$A,r_j,h_j,j$}
  \State $A$ redeems an $\mathsf{ExecuteDRT}$ bound to $(r_j,h_j)$
  \State $S$ burns the token and emits a redemption event carrying $(r_j,h_j)$
  \State $O$ resolves $r_j$ and forwards candidate computation $j$ to $E$ as $(r_j,h_j)$
  \State $E$ verifies $\mathsf{Hash}(j)=h_j$
  \State $E$ unseals the pool and executes $j$
  \State $E$ returns $j(\mathrm{pool})$ to $A$ over an encrypted channel
         \Comment{raw records never leave $E$}
\EndProcedure
\end{algorithmic}
\end{algorithm}

\begin{algorithm}[t]
\caption{P4: Settlement}
\label{alg:settlement}
\begin{algorithmic}[1]
\Require Analyst $A$, Solana program $S$, Enclave $E$, Analyst payment $p$

\Procedure{Settle}{$A,p$}
  \State $A$ receives $j(\mathrm{pool})$ from $E$
  \State $S$ distributes the royalty share of $p$ to the contributor-token holders
  \State $S$ records the redemption and computation identifier
\EndProcedure
\end{algorithmic}
\end{algorithm}

Finally, for \emph{P4} (Algorithm~\ref{alg:settlement}), the enclave returns the result to the analyst over an encrypted channel. A share of the analyst's payment is routed on-chain to the contributor-token holders pro rata. Revenue therefore follows contribution without requiring disclosure of which contributor supplied which record. The combination of (i)~on-chain provenance for every redemption and payout and (ii)~attested provenance for every enclave provides a tamper-evident audit trail of who computed what, on which dataset, and under which approved code.

\subsection{Data Pools}
\label{sec:pools}

A data pool is the abstraction that groups compatible contributions under a common schema and execution policy. A data pool is defined as follows:

\begin{definition}\label{Def::DataPool}
A \emph{data pool} is a data structure $\bigl(\sigma,\, \mathbf{r},\, \mathcal{C},\, K_p,\, \mathfrak{a}\bigr)$ where
\begin{itemize}
  \item $\sigma$ is a schema,
  \item $\mathbf{r}$ is a vector of permitted rights realised as DRT mints,
  \item $\mathcal{C}$ is an append-only multiset of contributor records governed by~$\sigma$,
  \item $K_p$ is a sealing key held only inside the pool's enclave, and
  \item $\mathfrak{a}$ is the Solana program-derived address (PDA) that anchors $\mathbf{r}$ to the pool.
\end{itemize}
\end{definition}

A pool is therefore a \emph{multi-contributor data set bound to a fixed schema and a fixed set of computations (DRTs)}, not a single user's upload. A creator opens a pool when there is value in making a class of records jointly analysable, and contributors then append records under the same schema. The $\mathsf{AppendDRT}$ mechanism enforces the attribution. Pools live in their own enclaves, and the trade-off is deliberate. Sealing keys are per-pool, so a compromise of one enclave cannot unseal records of another. The working set of an enclave is bounded by a single pool, which keeps memory pressure tractable, and the enclaves scale with the number of pools rather than with the volume of data. Pool creation is the only operation that requires a new enclave to be provisioned, and this is an infrequent operation amortised over the life of the pool since contributions and redemptions reuse the same enclave. Where pools are small or low-risk, multiple pools can be co-located in a single enclave by isolating them with per-pool sealing keys and schema-tagged interfaces. The prototype uses the simpler one-pool-per-enclave configuration. Contributors and analysts choose pools through the registry maintained on-chain by the smart contract. Each pool advertises its schema, its enclave measurement, the set of DRTs, and the redemption history.

\subsection{Governance: Resale, Permitted Computations, and Output}\label{sec:governance}

The code permitted by a DRT is fixed by the pool creator during pool creation (Algorithm~\ref{alg:pool-creation}) and is visible to all contributors and analysts. The creator therefore defines the allowable DRTs for the pool, including the associated code and terms (e.g., output policy). Three properties hold, under the platform and authorisation assumptions in Section~\ref{sec:threat-model}, regardless of operator behaviour. First, on resale, DRTs are freely transferable, but redeeming a DRT burns it, so the total number of times a computation can run on a pool equals the supply chosen during pool creation. Second, on permitted computations, each non-append DRT names an open-source program and a hash of its compiled artefact, and before execution the enclave fetches the program and refuses to run if the hash does not match. Third, on output controls, output is bounded per redemption by the open-source code referenced by each DRT and across redemptions by the DRT's burning supply. The pool creator therefore determines which computations may run on the pool, how many times they may run, and what form their outputs may take. For example, a creator may require aggregate-only outputs together with composable in-enclave wrappers such as differential-privacy noise or minimum-cohort guards. 

\subsection{Threat Model and Deployment Assumptions}
\label{sec:threat-model}

We separate trust into a \emph{platform layer} (the TEE and its remote attestation) and an \emph{authorisation layer} (the on-chain rules governing which code may run on which pool). At the platform layer we rely on standard TEE guarantees of code authenticity, runtime integrity, and confidentiality of data in use. We assume an untrusted host where infrastructure operators could in principle substitute malicious services without users noticing. To mitigate this, we use RA-TLS (Remote Attestation over TLS) to ensure that users only establish secure channels with genuine enclaves, and data at rest is protected by enclave-specific sealing keys. The prototype runs on Azure DCsv3~\cite{microsoft_dcsv3_azure} confidential VMs, but the design is TEE-agnostic, with AMD SEV-SNP~\cite{amd_sev_snp_2020}, Intel TDX~\cite{intel_tdx_2020}, and ARM CCA~\cite{arm_cca_2021} as viable alternatives. The authorisation layer rests on two assumptions. \textbf{(T1) Hardware:} the TEE faithfully executes the loaded image and remote attestation binds its measurement to a verifiable quote. We accept the published adversary model for the underlying TEE family and treat physical attacks and microarchitectural side channels~\cite{VanBulck2018,Chen2019,Murdock2020} as out of scope. \textbf{(T2) Authorisation:} pool creators choose, and contributors accept, a set of computations (DRTs) whose outputs they are willing to release. The DRTs are public, immutable, and bound to on-chain code hashes. This split clarifies that the residual risk of authorising malicious code (DRT) is a T2 concern rather than a T1 failure---the hardware proves which code ran---governance decides whether that code should have run. Permitted computations (DRTs) are open-source Python scripts or WASM binaries that users can verify via hashes. The DRT redemptions are recorded on-chain and relayed to enclaves by oracle nodes (we start with a single oracle node, but the design supports $k$-of-$n$ consensus for stronger guarantees).

Our platform prioritises control rather than secrecy, as computation results are returned. Thus, users cede some privacy, but the chosen computation (DRT) limits the scope of leakage and to whom. Ensuring that only approved code runs is the key challenge, as attestation and hash verification constrain execution. Given the known limitations of TEEs, we treat them as an enabler rather than a solution in themselves. Our contribution lies in coupling attested compute with on-chain rights to operationalise enforceable \emph{control}. Concretely, this requires a layered stance: (i) minimise the trusted computing base with narrow enclave interfaces; (ii) bind code and configuration to attestation, require signed updates, and block rollbacks; (iii) constrain outputs via aggregation, rate limiting, and computation-specific caps; and (iv) harden operations with audit logs, deny/allow lists of measurements, and eventually multi-oracle consensus. The novelty lies in integrating these layers so that data creators retain ex ante, enforceable control over how their data is processed. Appendix~\ref{sec:challenges} discusses recurring engineering challenges, and Appendix~\ref{sec:output-resale} addresses output resale and compositional reconstruction.

\section{Feasibility and Preliminary Evaluation}

Our implementation demonstrates the end-to-end feasibility of the proposed information infrastructure. Specifically, it supports the creation of data pools, the issuance and redemption of DRTs as Solana SPL tokens, and the execution of WASM and Python jobs over sealed datasets using Intel SGX. Our prototype demonstrates that the core workflows can be realised in practice: pool instantiation, token redemption, code verification, and in-enclave execution. 
We present the prototype as evidence that verifiable, control-centric computation is technically feasible, and we hope this forms a foundation for further evaluation. A key inhibitor in the current implementation is the time required to initialise a new enclave for each data pool, although this is assumed to be an infrequent operation. Moreover, throughput under varying pool sizes requires further investigation, as our initial tests rely on small datasets and simplistic DRT-authorised code (e.g., computing mean or median values).  

Our work connects to several lines of research. First, privacy-preserving data sharing has traditionally relied on cryptographic primitives such as secure multiparty computation~\cite{Yao1986}, homomorphic encryption~\cite{Gentry2009}, or frameworks such as differential privacy~\cite{Dwork2006}. However, these approaches prioritise secrecy but often suffer from performance or usability limitations. Second, TEEs such as Intel SGX~\cite{Anati2013,Costan2016IntelSGX} have been applied to protect sensitive computations, though prior work highlights their side-channel vulnerabilities~\cite{VanBulck2018,Chen2019,Murdock2020}. Third, tokenised data governance has been explored in blockchain-based data markets~\cite{Missier2017,Xiao2020}, but such systems typically emphasise access rights, licensing, or accountability without enforceable control over what computations are performed on the underlying data. Our contribution lies in integrating these strands by binding narrowly scoped digital rights (DRTs) to auditable code and attested TEEs, thereby operationalising privacy as \emph{control over use} rather than secrecy alone. 
\bibliographystyle{splncs04}
\bibliography{bibliography}
%







\appendix
\renewcommand{\theHsection}{appendix.\Alph{section}}

\section{Engineering Challenges}
\label{sec:challenges}

Although the system combines familiar building blocks (a TEE, a smart contract, an oracle, and a frontend), prototyping surfaced five non-trivial challenges that we expect to recur in any deployment of this pattern. 

\paragraph{C1: Binding on-chain rights to off-chain execution.} The DRT contract is on-chain and the enclave is off-chain. The link between them must satisfy both safety and liveness. Safety requires that one redemption authorises at most one execution, even if the same event is forwarded multiple times. Liveness requires that a valid redemption eventually reaches the enclave. Our prototype uses a single oracle, so a malicious or faulty oracle can delay, censor, or omit an execution. However, it should not cause a correctly implemented enclave to execute an unauthorised computation, provided that forwarded redemptions are checked against authenticated ledger state rather than trusted as oracle assertions. Because redemptions are durably recorded on-chain, missed events can be recovered and reprocessed from ledger history. However, putting a full RPC client, Solana verifier, or similar blockchain logic inside the enclave would increase the trusted computing base and is therefore undesirable. The enclave should remain minimal and only verify the smallest necessary evidence. Threshold-based oracle quorums can improve liveness and reduce reliance on a single relay, but require deduplication of forwarded events inside the enclave.

\paragraph{C2: Reconciling code identity.} A program admitted to a pool's DRT library carries two identities, namely the source repository (e.g., a GitHub link) and the cryptographic hash of the program. Any change to the source, however minor, will produce a different hash and invalidate the corresponding DRT until the source is reverted. This rigidity is a feature rather than a bug, since it guarantees that what an analyst redeems is exactly what was admitted, but it places the burden of versioning on pool creators. In practice, code creators can publish revised programs as new DRTs alongside the originals rather than overwriting them.  We leave more sophisticated approaches, such as semantic versioning bound to attestation policies, to future work.

\paragraph{C3: Operational hygiene of attestation.} Attestation evidence is not write-once. Collateral expires, signing keys rotate, and TEE microcode receives advisories. A ``provision once, run forever'' flow breaks under any of these. The deployment service should re-attest enclaves on every restart and on a daily schedule, and reject enclaves whose most recent quote is stale. This is a small operational detail with practical consequences for any long-lived confidential-computing deployment. Deterministic builds of the underlying SDK (used per enclave) are also an essential requirement for attestation.

\paragraph{C4: Schema evolution.}  Once a pool is sealed under a schema, that schema is effectively immutable. Changing it would either invalidate sealed records or require a re-encryption step that exposes the records to whoever performs it. The prototype handles migration by opening a new pool under the revised schema and inviting contributors to re-deposit, which preserves the integrity of the original pool but loses the historical contributor base unless every member re-engages. This is workable for the data classes we target but genuinely costly at scale, and we flag it as an open problem.

\paragraph{C5: Authorisation of programs.} Even with hash-bound execution, a contributor can be socially engineered into joining a pool whose DRTs contain a malicious program. Hardware-based attestation can prove which program ran (i.e., the underlying SDK runtime), but it cannot decide whether the program referenced by the DRT is benign. The DRT library makes this risk visible at admission rather than at execution, although eliminating it lies outside the remit of any single protocol layer (Section~\ref{sec:threat-model}, T2).

\section{Output Resale and Composition Risk}
\label{sec:output-resale}

\paragraph{Output resale.} Once a DRT is redeemed and a result is returned, the analyst possesses a digital artefact that the protocol cannot itself prevent from being copied or resold (i.e., the computation result). This is intrinsic to control-centric privacy: we govern \emph{computation}, not the downstream handling of derived outputs. Two mechanisms mitigate the impact. First, supply-bounded DRTs cap the number of times any computation runs against a pool, so resale of a result does not enable unbounded re-execution. Second, output policies (DRTs) chosen by the pool creator bound the value of a single result to a would-be reseller. 

\paragraph{Compositional reconstruction.}

A pool creator who admits a sufficiently expressive set of DRTs may enable reconstruction of individual records by combining outputs across redemptions. Our design surfaces this risk at admission by making the full DRT set, code references, and supplies public on-chain before any contribution. Creators are therefore responsible for choosing computations whose joint output channel is acceptable, and contributors observe this set before redeeming an $\mathsf{AppendDRT}$. Formal composition analysis (e.g., differential-privacy budgets across DRTs, or query-set auditing) is a natural extension but lies outside the scope of this paper.

\end{document}